\documentclass[12pt,thmsa]{article}
\usepackage{amssymb}
\usepackage{amsfonts}
\usepackage{sw20aip}
\usepackage[T1]{fontenc} 

\input{tcilatex}
\begin{document}

\author{Carlos Kozameh$^{1}$ and Ezra T. Newman$^{2}$ \\
$^{1}$FaMaF, Univ. of Cordoba, \\
Cordoba, Argentina\\
$^{2}$Dept of Physics and Astronomy, \\
University of Pittsburgh,\\
Pittsburgh, PA 15260. USA}
\title{Asymptotically Shear-free and Twist-free Null Geodesic Congruences }
\date{March 20, 2007 }
\maketitle

\begin{abstract}
We show that, though they are rare, there are asymptotically flat
space-times that possess null geodesic congruences that are both
asymptotically shear- free and twist-free (surface forming). In particular,
we display the class of space-times that possess this property and
demonstrate how these congruences can be found. A special case within this
class are the Robinson- Trautman space-times. In addition, we show that in
each case the congruence is isolated in the sense that there are no other
neighboring congruences with this dual property.
\end{abstract}

\section{\protect\bigskip Introduction}

For an arbitrary asymptotically flat space-time, in general, it is easy to
find families of null geodesic congruences that are \textit{either}
twist-free or are asymptotically shear-free. To find congruences with both
properties: i.e., a null geodesic congruences that is simultaneously
twist-free (i.e., surface forming) and asymptotically shear-free is
difficult and often impossible. By taking an arbitrary one parameter family
of slicing of null infinity, $\mathfrak{I}^{+}$, the inward directed null
normals generate a twist-free null geodesic congruences. On the other hand,
the asymptotically shear-free null geodesic congruences, (which are obtained
from any shearing twist-free Bondi congruence via solutions of the good cut
equation\cite{footprints}), almost always contain twist. It has proven
difficult to find examples of asymptotically flat space-times where one can
find congruences that are both asymptotically shear-free and twist-free.
They certainly are not generic. Special examples of such space-times are
given by the class of regular Robinson-Trautman metrics, where the
congruence is given by the twist-free degenerate principle null vectors of
the Weyl tensor. It turns out that such space-times and their associated
congruences are of considerable interest for studies of certain issues in
the asymptotic behavior of gravitational fields\cite{footprints,RT,RTM}. It
is the purpose of this note to show how large classes of such space-times
can be constructed along with their associated asymptotically shear-free and
twist-free congruence. In a future paper we will investigate their physical
properties.

\section{Construction}

The basic problem is how to construct on $\mathfrak{I}^{+},$(with given
Bondi coordinates\textbf{\ (}$u,\zeta ,\overline{\zeta },r$\textbf{)},
special families of Bondi asymptotic shear, $\sigma (u,\zeta ,\overline{%
\zeta }),$ such that one can find from them\textit{\ a one parameter family
of real} solutions to the so-called `good cut' equation, 
\begin{equation}
\text{\dh }^{2}Z=\sigma (Z,\zeta ,\overline{\zeta }).  \label{good cut}
\end{equation}%
In general, the solutions,\textbf{\ }$Z(z^{a},\zeta ,\overline{\zeta }),$ 
\textbf{\ }of the good cut equation\ depend on four complex constants, 
\textbf{\ }$z^{a},$\textbf{\ }\{which define H-space\}. For each arbitrarily
chosen complex curve on the H-space, i.e., for $z^{a}=\xi ^{a}(\tau ),$ with
complex parameter $\tau ,$ one can easily construct an asymptotically
shear-free null geodesic congruence\cite{footprints}. This is done by taking
the solution, 
\begin{equation}
u=X(\tau ,\zeta ,\overline{\zeta })\equiv Z(\xi ^{a}(\tau ),\zeta ,\overline{%
\zeta })  \label{NUComplex}
\end{equation}%
and then defining (parametrically) the complex stereographic angle field, $%
L(u,\zeta ,\overline{\zeta })$ on $\mathfrak{I}^{+},$ by 
\begin{eqnarray}
L(u,\zeta ,\overline{\zeta }) &=&\text{\dh }_{(\tau )}X(\tau ,\zeta ,%
\overline{\zeta })  \label{L} \\
u &=&X(\tau ,\zeta ,\overline{\zeta })\text{ }{\large \Leftrightarrow }\text{
}\tau =T(u,\zeta ,\overline{\zeta }).  \label{T}
\end{eqnarray}%
(\dh $_{(\tau )}$ denotes the \dh\ operator holding $\tau $ constant.) The
shear-free congruence, defined and determined by the asymptotic vector 
\textbf{\ }$l^{\ast a},$ is then obtained from the Bondi congruence by the
null rotation, near $\mathfrak{I}^{+},$ of their respective tangent vectors,
($l^{\ast a},l_{B}^{a}$ ) by

\begin{eqnarray}
l^{\ast a} &=&l_{B}^{a}-\frac{L}{r}\overline{m}_{B}^{a}+\frac{\overline{L}}{r%
}m_{B}^{a}+O(r^{-2}),  \label{NullRot} \\
m^{\ast a} &=&m_{B}^{a}-\frac{L}{r}n_{B}^{a}+O(r^{-2}),  \nonumber \\
\overline{m}^{\ast a} &=&\overline{m}_{B}^{a}-\frac{\overline{L}}{r}%
n_{B}^{a}+O(r^{-2}),  \nonumber \\
n^{\ast a} &=&n_{B}^{a},  \nonumber
\end{eqnarray}%
where ($l_{B}^{a},n_{B}^{a},m_{B}^{a},\overline{m}_{B}^{a}$) form the Bondi
null tetrad system at $\mathfrak{I}^{+}.$ The twist, $\Sigma ,$ of $l^{\ast
a}$ is given by 
\begin{equation}
i\Sigma \equiv \frac{1}{2}\{\text{\dh }\overline{L}+L\overline{L}^{{\large %
\cdot }}-\overline{\text{\dh }}L-\overline{L}L^{{\large \cdot }}\}.
\label{SIGMA}
\end{equation}%
where dot denotes u-derivative.

Our desire is to be able to chose real world-lines $z^{a}=\xi ^{a}(s),$ $%
\tau =s,$ so that the solutions to Eq.(\ref{good cut}) take the form 
\begin{equation}
u=X(s,\zeta ,\overline{\zeta })  \label{NU}
\end{equation}
with both $s$ and $X$ real$.$ Such solutions, if they could be found, define
a (real) slicing (or NU coordinates) of $\mathfrak{I}^{+},$ by $s=const.,$
where the null inward normals form null surfaces, i.e. are twist-free and
are asymptotically shear-free. This later result follows from the fact that
Eq.(\ref{NU}) is a solution to the 'good cut equation'\cite{footprints}. The
difficulty has been that Eq.(\ref{good cut}) is a complex differential
equation and almost all solutions are `intrinsically' complex. It was not at
all clear what is the class of Bondi shears that allow for these
one-parameter families of real solutions.

The first thing to note is that for a completely general Bondi shear there
are two real functions of $(u,\zeta ,\overline{\zeta })$ that are buried in
the (complex) shear, $\sigma (u,\zeta ,\overline{\zeta }),$ and we want one
real function, $u=X(s,\zeta ,\overline{\zeta }),$ to solve the equation - or
from another point of view\cite{A}, to `transform the shear away'. \textbf{\ 
}We look for a Bondi shear that depends on only one real function.

The construction of appropriate Bondi shears, though not immediately
obvious, turns out to be quite simple. We begin by choosing an arbitrary 
\textit{real} function of three variables, 
\begin{equation}
u=H(s,\zeta ,\overline{\zeta }).  \label{H}
\end{equation}
that can be interpreted as yielding a one-parameter slicing of $\mathfrak{I}%
^{+}. $ The inward directed null normals, as we mentioned earlier, form a
null surface, i.e., are twist free by construction. Our task is to construct
asymptotically flat space-times possessing a special class of Bondi shears
so that the asymptotic shear of these normal null rays vanishes.

We assume that Eq.(\ref{H}) can be inverted as 
\begin{equation}
s=S(u,\zeta ,\overline{\zeta }).  \label{NU Inverse}
\end{equation}
Next (by applying \dh $^{2}$ to $H$, holding $s$ constant) we define the
spin-wt-2 complex `electric type' function 
\begin{equation}
\sigma ^{*}(s,\zeta ,\overline{\zeta })\equiv \text{\dh }_{(s)}^{2}H(s,\zeta
,\overline{\zeta }).  \label{sigma**}
\end{equation}
Finally the Bondi shear is obtained by replacing the $s$ in Eq.(\ref{sigma**}
) by $s=S(u,\zeta ,\overline{\zeta })$ obtaining 
\begin{equation}
\sigma (u,\zeta ,\overline{\zeta })\equiv \sigma ^{*}(S(u,\zeta ,\overline{%
\zeta }),\zeta ,\overline{\zeta })  \label{spec.shear}
\end{equation}
so that the good cut equation becomes 
\begin{equation}
\text{\dh }^{2}X=\sigma (X,\zeta ,\overline{\zeta })\equiv \text{\dh }%
_{(s)}^{2}H(S(X,\zeta ,\overline{\zeta }),\zeta ,\overline{\zeta }).
\label{nice good cut eq}
\end{equation}

We easily see, from the construction and Eq.(\ref{sigma**}), that a real
solution is given by 
\begin{equation}
u=X(s,\zeta ,\overline{\zeta })=H(s,\zeta ,\overline{\zeta }).
\label{RealSolution}
\end{equation}

We thus see that for any choice of the function $u=H(s,\zeta ,\overline{%
\zeta }),$ (an NU coordinate transformation), we can define or construct a
Bondi shear on the space-time so that the null normals to the constant `$s$'
cuts of $\mathfrak{I}^{+}$ form a congruence that is both asymptotically
shear-free and twist-free. The $l=0,1$ harmonic components of $H,$ written
as $\xi ^{a}(s)l_{a}(\zeta ,\overline{\zeta }),$ determine the associated
real world line, $\xi ^{a}(s)$. From the construction, it is clear that
this, Eq.(\ref{spec.shear}), is the only class of Bondi shears with this
dual property.

\begin{remark}
In this construction it might appear that by simply adding to the H in Eq.( %
\ref{RealSolution}) another term of the form $\xi ^{\#a}(s)l_{a}(\zeta ,%
\overline{\zeta })$ a new solution could be constructed. This is not true
since changing H by that term changes the $S(u,\zeta ,\overline{\zeta })$
and thus the Bondi shear itself. The associated space-time would also
change. However, if we are given the modified special shear from Eq.(\ref%
{spec.shear}) 
\begin{equation}
\sigma (u,\zeta ,\overline{\zeta })\equiv \sigma ^{\ast }(S(u,\zeta ,%
\overline{\zeta }),\zeta ,\overline{\zeta })+\text{\dh }^{2}\alpha (\zeta ,%
\overline{\zeta })
\end{equation}%
(which defines an arbitrary spin.wt.2 function, $F(\zeta ,\overline{\zeta })=
$ \dh $^{2}\alpha (\zeta ,\overline{\zeta }))$ the extra term can be
considered as pure gauge since the good cut equation, \dh $^{2}X=\sigma
^{\ast }(X,\zeta ,\overline{\zeta })+$\dh $^{2}\alpha (\zeta ,\overline{%
\zeta }),$ \ becomes%
\[
\eth ^{2}(X-\alpha (\zeta ,\overline{\zeta })\equiv \eth ^{2}X^{\ast
}=\sigma ^{\ast }(X,\zeta ,\overline{\zeta })
\]%
so that the gauge choice is equivalent tothe choice of a Bondi
supertranslation.
\end{remark}

Another question that immediately arises: Given the Bondi shear, Eq.(\ref%
{spec.shear}), and the real one-parameter family of solutions,\textbf{\ }%
Eq.( \ref{RealSolution}), are there other \textit{real solutions }close to
the first real solution?

For the generic situation the answer is no.

To show this we begin with the good cut equation 
\[
\text{\dh }^{2}Z=\sigma (Z\text{,}\zeta \text{,}\overline{\zeta }) 
\]
and assume that the shear is of the form Eq.(\ref{spec.shear}) and that we
are considering the \textit{real} one parameter solution and the nearby
solutions. We know from the general theory\cite{HSpace1,HSpace2} that there
is a complex four parameter family of solutions, $u=Z(z^{a},\zeta ,\overline{%
\zeta })$ and hence the \textit{real} one parameter solution has an
associated real world-line, $z^{a}=\xi ^{a}(s),$ with 
\begin{equation}
X(s,\zeta ,\overline{\zeta })=Z(\xi ^{a}(s),\zeta ,\overline{\zeta }).
\label{real X}
\end{equation}

For a \textit{given} real point on the real world-line, $z^{a}=\xi
^{a}(s_{0}),$ we consider \textit{real neighboring points}, $z^{a}=\xi
^{a}(s_{0})+\delta \xi ^{a}.$ More specifically we consider two different
neighboring points; one on the world-line, 
\begin{equation}
z^{a}=\xi ^{a}(s_{0})+\xi ^{a\prime }(s_{0})ds  \label{online*}
\end{equation}
the other off the world-line, 
\begin{equation}
z^{a}=\xi ^{a}(s_{0})+\eta ^{a}(s_{0})\epsilon  \label{online}
\end{equation}
with 
\begin{equation}
\eta ^{a}(s_{0})\neq \mu \xi ^{a\prime }(s_{0}).  \label{different*}
\end{equation}

Given the solution, $u=X(s_{0},\zeta ,\overline{\zeta })=Z(\xi
^{a}(s_{0}),\zeta ,\overline{\zeta })\equiv Z_{0},$ neighboring solutions, 
\[
Z(\xi ^{a}+\delta \xi ^{a},\zeta ,\overline{\zeta })=Z_{0}+\partial
_{a}Z_{0}\delta \xi ^{a} 
\]%
are determined by the solutions, $\partial _{a}Z_{0},$ to the linearized
equation 
\begin{equation}
\text{\dh }^{2}(\partial _{a}Z_{0})=\dot{\sigma}\text{(}Z_{0}\text{,}\zeta ,%
\overline{\zeta }\text{)}\cdot \partial _{a}Z_{0}.  \label{gradZ}
\end{equation}%
The fact that we already know one solution, namely the one coming from
displacement along the known world-line, Eq.(\ref{online*}), i.e., 
\begin{equation}
V=\partial _{a}Z_{0}\cdot \xi ^{a\prime }(s_{0})\equiv \partial
_{s}X(s,\zeta ,\overline{\zeta })  \label{V}
\end{equation}%
allows us the find the general solution. Prime denotes s-derivative.

Assuming the general (four-parameter) solution has the form 
\begin{equation}
\partial _{a}Z_{0}=VL_{a},  \label{test}
\end{equation}%
Eq.(\ref{gradZ}) becomes 
\begin{equation}
\text{\dh }^{2}(VL_{a})\equiv \text{\dh }^{2}V\cdot L_{a}+2\text{\dh }V\cdot 
\text{\dh }L_{a}+V\text{\dh }^{2}L_{a}=\dot{\sigma}VL_{a}
\end{equation}%
or 
\begin{eqnarray}
2\text{\dh }V\text{\dh }L_{a}+V\text{\dh }^{2}L_{a} &=&0 \\
\text{\dh (}V^{2}\text{\dh }L_{a}) &=&0  \nonumber
\end{eqnarray}%
with solution 
\begin{eqnarray}
V^{2}\text{\dh }L_{a} &=&\gamma _{a}^{i}m_{i}  \label{solution} \\
m_{i} &=&\frac{\sqrt{2}}{2P}(1-\overline{\zeta }^{2},-i(1+\overline{\zeta }%
^{2}),\text{ }2\overline{\zeta }) \\
\gamma _{a}^{i} &=&\text{three constant `vectors'}
\end{eqnarray}

Assuming (from genericity) that $V^{-2}$ contains a non-vanishing $l=0$
spherical harmonic component, say $K,$ we can write 
\begin{equation}
V^{-2}=K+\text{\dh }W  \label{V^-2}
\end{equation}
with $W$ a spin.wt.(-1) quantity, (containing, in general, all harmonics, $%
l\geq 1$). Using Eq.(\ref{V^-2}), Eq.(\ref{solution}) can be rewritten as 
\begin{eqnarray*}
\text{\dh }L_{a} &=&V^{-2}\gamma _{a}^{i}m_{i} \\
&=&\gamma _{a}^{i}(K+\text{\dh }W)m_{i} \\
&=&\gamma _{a}^{i}(Km_{i}+m_{i}\text{\dh }W) \\
&=&\gamma _{a}^{i}(K\text{\dh }l_{i}+\text{\dh }[Wm_{i}]) \\
&=&\text{\dh }\{\gamma _{a}^{i}(Kl_{i}+Wm_{i})\} \\
l_{i} &=&\frac{\sqrt{2}}{2P}(\zeta +\overline{\zeta },\text{ }-i(\zeta -%
\overline{\zeta }),-1+\zeta \overline{\zeta })
\end{eqnarray*}
or 
\[
L_{a}=\gamma _{a}^{0}+\gamma _{a}^{i}(Kl_{i}+Wm_{i}) 
\]
with $\gamma _{a}^{0}$ another independent constant `vector'.

The general solution, i.e., Eq.(\ref{test}), becomes

\begin{equation}
\partial _{a}Z_{0}=V\{\gamma _{a}^{0}+\gamma _{a}^{i}(Kl_{i}+Wm_{i})\}.
\label{test*}
\end{equation}
The four constant `vectors', ($\gamma _{a}^{0},\gamma _{a}^{i}$), are
determined (partially) by the condition that$,$ 
\begin{equation}
V\equiv \partial _{a}Z_{0}\xi ^{a\prime }=V\{\xi ^{a\prime }\gamma
_{a}^{0}+\xi ^{a\prime }\gamma _{a}^{i}(Kl_{i}+Wm_{i})\}  \label{V=V}
\end{equation}
or

\[
\xi ^{a\prime }\gamma _{a}^{0}+\xi ^{a\prime }\gamma
_{a}^{i}(Kl_{i}+Wm_{i})\}=1. 
\]
Since ($l_{i},W,m_{i})$ have angular behavior the four `vectors' must
satisfy the algebraic conditions 
\begin{eqnarray}
\xi ^{a\prime }\gamma _{a}^{0} &=&1  \label{1} \\
\xi ^{a\prime }\gamma _{a}^{i} &=&0.  \label{2}
\end{eqnarray}

Assuming that we have found a set of four independent `vectors' with the
given $\xi ^{a\prime }$ that satisfy Eqs.(\ref{1}) and (\ref{2}), then 
\textit{for any other real quantity, }$\eta ^{a}(s_{0}),$with\textit{\ } 
\begin{equation}
\eta ^{a}(s_{0})\neq \mu \xi ^{a\prime }(s_{0})  \label{not equal}
\end{equation}
we have that 
\[
\eta ^{a}(s_{0})\gamma _{a}^{i}\neq 0. 
\]
This immediately implies that \textit{all solutions} that differ from $V$ or
a multiple of $V,$ i.e., 
\[
\partial _{a}Z_{0}\eta ^{a}=V\{\eta ^{a}\gamma _{a}^{0}+\eta ^{a}\gamma
_{a}^{i}(Kl_{i}+Wm_{i})\} 
\]
are intrinsically complex, so that in the generic situation there are no
other real solutions in the neighborhood of Eq.(\ref{RealSolution}).

\section{Conclusion}

We studied here the question of: what is the class of space-times that
possess null geodesic congruences that are both asymptotically shear-free
and twist-free and furthermore asked, within that class how common are such
congruences?

Our motivation for this work was the recognition that the well-known
Robinson-Trautman metrics allow such a congruence - and in fact are
essentially defined by a special case in this class, namely the ones that
are globally shear-free and twist-free and not just asymptotically
shear-free and twist-free. This work can thus be considered as attempting to
find a generalization of the Robinson-Trautman case.

In a future paper we will apply the results of this work to study the
relationship of the \textit{real world-line} to the Bondi energy-momentum
four-vector and the resulting `equations of motion'\cite{RTM} coming from
the asymptotic Einstein-Maxwell equations, i.e., those coming from the Bondi
energy-momentum loss equation. Though it might be hard to believe, we are
convinced that this future study can and will shed light on the run-away
behavior associated with the radiation reaction term in the electrodynamic
equations of motion for a charged particle\cite{RTM}.

\section{Acknowledgments}

This material is based upon work (partially) supported by the National
Science Foundation under Grant No. PHY-0244513. Any opinions, findings, and
conclusions or recommendations expressed in this material are those of the
authors and do not necessarily reflect the views of the National Science.
E.T.N. thanks the NSF for this support. C.K. thanks CONICET and SECYTUNC for
support. \ In addition we thank our colleague, Gilberto Silva-Ortigoza, for
comments and suggestions.

\end{document}